\documentclass[twocolumn,showpacs,amsmath,amssymb,prl,superscriptaddress]{revtex4-1}
\usepackage{graphicx}
\usepackage{epstopdf}
\usepackage{epsfig}
\usepackage{dcolumn}
\usepackage{color}
\usepackage{bm}
\usepackage{mathrsfs}
\usepackage{bbold}
\usepackage{amssymb}
\usepackage{upgreek}

\begin{document}

\title{Nonequilibrium transient dynamics of photon statistics}

\author{Md.~Manirul Ali}
\email{mani@mail.ncku.edu.tw}
\affiliation{Department of Physics, National Cheng Kung University, Tainan 70101, Taiwan}
\author{Wei-Min Zhang}
\email{wzhang@mail.ncku.edu.tw}
\affiliation{Department of Physics, National Cheng Kung University, Tainan 70101, Taiwan}

\date{\today}

\begin{abstract}

We investigate the transient dynamics of photon statistics through two-time correlation functions for
optical fields. We find that the {\it transient} correlations at different time $t$ yield a smooth transition from
antibunching to bunching photon statistics in the weak system-environment coupling regime. In the strong-coupling
regime, the two-time correlations exhibit bunching-antibunching oscillations that persists
both in the transient process and in the steady-state limit. The photon bunching-antibunching oscillations
is a manifestation of strong {\it non-Markovian} dynamics, where the system remains in nonequilibrium
from its environment. We also find that the antibunching to bunching transition in the weak-coupling regime and the
bunching-antibunching oscillation in the strong-coupling regime are strongly influenced by the initial environment
temperature.

\end{abstract}

\maketitle

Photon quantum statistics, such as photon {\it antibunching}, has fundamental importance in understanding, generating
and manipulating the nonclassical states of light applicable in quantum optics and quantum information processing. Photon
{\it bunching} \cite{HBT1} is the tendency of photons
to distribute themselves in bunches without having any time delay between them. Photon antibunching refers to the statistical property
of a light field where the probability of time delayed photon increases \cite{Scully97}. Photon bunching and antibunching statistics are
usually characterized by the steady-state
second-order correlation function $g^{(2)}_{ss}(\tau)$, where an increasing (decreasing) magnitude of
$g^{(2)}_{ss}(\tau)$ with delay-time $\tau$ demonstrate antibunching (bunching) statistics of photons. Photon antibunching was first
observed in resonance fluorescence, which was also the first observed nonclassical effect requiring a full quantum description
of light \cite{Kimble77}. Since then, such quantum effects have been experimentally explored in strong nonlinear systems,
such as an optical cavity strongly coupled to trapped atoms \cite{cavityQED1,cavityQED2}, emitted photons from a single quantum dot
at room temperature \cite{QDot1}, quantum dot coupled to photonic crystal resonator \cite{QDot2}, superconducting qubit coupled
to a microwave cavity \cite{SCQ1}, fluorescence from nitrogen-vacancy center in diamond \cite{NVCenter}, and in coupled
optomechanical systems \cite{OMS1,OMS2}. In all these investigations, antibunching is manifested mainly for stationary field
through $g^{(2)}(\tau)$. In this Letter, we will explore nonequilibrium transient dynamics of photon statistics through the
second-order correlation function $g^{(2)}\left(t, t+\tau \right)$.

The transient second-order correlation function $g^{(2)}\left(t, t+\tau \right)$ is explicitly determined by the
two-time correlation function $\langle a^{\dagger}(t) a^{\dagger}(t+\tau) a(t+\tau) a(t) \rangle$ as
\begin{eqnarray}
g^{(2)}\left(t, t+\tau \right) = \frac{\langle a^{\dagger}(t) a^{\dagger}(t+\tau) a(t+\tau) a(t) \rangle}
{\langle a^{\dagger}(t) a(t) \rangle \langle a^{\dagger}(t+\tau) a(t+\tau) \rangle},
\label{g2}
\end{eqnarray}
which is directly related to the correlations between two photons, one detected at time $t$ and another at time $t+\tau$.
The steady-state correlation function $g^{(2)}_{ss}(\tau)$ measured in above experiments \cite{Kimble77,cavityQED1,cavityQED2,QDot1,QDot2,SCQ1,NVCenter,OMS1,OMS2} is the long time steady-state
limit of $g^{(2)}\left(t, t+\tau \right)$ as
$t \rightarrow \infty$. The transient dynamics of the second-order correlation function $g^{(2)}\left(t, t+\tau \right)$ at
arbitrary $t$ have special significance in understanding nonequilibrium dynamics. They provide important information
of non-Markovian back-action memory effects \cite{Fleming1,Fleming2}. The transient two-time correlation function
also plays a crucial role in the dissipative dynamics of many-body quantum systems \cite{ManyBody1,ManyBody3}
and transient quantum transport dynamics when the system is out-of-equilibrium \cite{Transient1,Transient2,Transient3}.
In this Letter, we find that the transient $g^{(2)}\left(t, t+\tau \right)$ exhibits various transitions between bunching and
antibunching which is hitherto unexplored.

We consider an optical field interacting with a thermal environment, modeled as a collection of infinite modes.
This system has been described by the famous Fano-model \cite{Fano} that has wide applications in atomic, photonic
and condensed matter physics \cite{photonR,Mahan}. In the previous investigations, the two-time correlation
function $\langle a^{\dagger}(t) a^{\dagger}(t+\tau) a(t+\tau) a(t) \rangle$ is usually calculated through
{\it quantum regression theorem} using Born-Markov approximation \cite{Scully97}
\begin{eqnarray}
\nonumber
&&\!\!\!\!\!\! \langle a^{\dagger}(t) a^{\dagger}(t+\tau) a(t+\tau) a(t) \rangle_{M} \\
\nonumber
&\!\!\!=& \left[\alpha e^{- 4 \kappa t} + 2{\bar n}(1- e^{- 2 \kappa t}) \{ (2\beta - {\bar n})e^{- 2 \kappa t} + {\bar n} \}  \right] e^{- 2 \kappa \tau} \\
&&{} + {\bar n} \left[ \beta e^{- 2 \kappa t} + {\bar n} (1 - e^{- 2 \kappa t}) \right] (1 - e^{- 2 \kappa \tau}),
\label{g2M}
\end{eqnarray}
where $\alpha\!=\!\langle a^{\dagger}(0) a^{\dagger}(0) a(0) a(0) \rangle$, $\beta\!=\!\langle a^{\dagger}(0) a(0) \rangle$, 
$\kappa$ is related to the dissipation of the optical field, and ${\bar n}$ is the average thermal photon number. Under such an
approximation, the steady-state limit of $g^{(2)}(t,t+\tau)$ is given by Eq.~(\ref{g2})
\begin{eqnarray}
g^{(2)}_{ss}(\tau) = \lim_{t \rightarrow \infty} g^{(2)}(t,t+\tau) = 1 + e^{- 2 \kappa \tau}.
\end{eqnarray}
which explains the well known Hanbury-Brown-Twiss effect or photon bunching for thermal light \cite{HBT1}.
By taking the Markov limit, it essentially ignores the transient dynamics of the correlation function. Whereas many quantum optics
devices exhibit {\it non-Markovian} memory effect for which the Born-Markov approximation is not applicable. Here
we show a nontrivial transient dynamics of $g^{(2)}(t, t+ \tau)$ in connection to photon bunching and antibunching
when the optical field interacts with a general non-Markovian environment. By solving the exact quantum Langevin
equation \cite{Tan,Transient2}, we can obtain the exact time evolution of
the optical field operator $a(t)$. Then, the two-time correlation function is given by
\begin{eqnarray}
\label{2order}
&&\langle a^{\dagger}(t) a^{\dagger}(t^{\prime}) a(t^{\prime}) a(t) \rangle \\
\nonumber
&&{} =  v(t) v(t^{\prime}) + \left|v(t,t^{\prime})\right|^2 + \left|u(t)\right|^2 \left|u(t^{\prime})\right|^2 \alpha \\
\nonumber
&&{} \!+\! \{ v(t) \left|u(t^{\prime})\right|^2 \!+\! v(t^{\prime}) \left| u(t) \right|^2  \!+\!
2 \textrm{Re} \left[ v(t,t^{\prime}) u^{\ast}(t) u(t^{\prime}) \right] \} \beta
\end{eqnarray}
which is determined by two basic Green's functions $u(t,0)=\langle [a(t),a^{\dagger}(0)] \rangle$ and
$v(t,t^{\prime})=\langle a^{\dagger}(t^{\prime}) a(t) \rangle$ (plus an initial state dependent term)
in {\it nonequilibrium} quantum systems \cite{general,kadanoff,Schwinger,Keldysh}.
We have denoted $u(t)\!=\!u(t,0)$, $v(t)\!=\!v(t,t)$, and $\alpha$, $\beta$ are already given after Eq.~(\ref{g2M}).
The nonequilibrium Green's function $u(t,0)$ satisfies the following integro-differential equation
\begin{eqnarray}
{\dot u}(t,0) + i \omega_0 u(t,0) + \int_{0}^t d\tau g(t,\tau) u(\tau,0) = 0,
\label{ide}
\end{eqnarray}
where $\omega_0$ is the frequency of the optical field. The integral kernel $g(t,\tau)$ describes the non-Markovian
back-action between the system and the environment, and can be determined by the spectral density
$J(\omega)$ through the relations: $g(t,\tau) = \int_0^{\infty} d\omega  J(\omega) e^{-i\omega(t-\tau)}$. The spectral
density of the environment is defined by $J(\omega)= \sum_k |V_k|^2 \delta(\omega-\omega_k)$, where
$V_k$ specifies the coupling between the system and the $k$-th mode of the environment. The correlation Green's function
$v(t,t^{\prime})$ which characterizes the nonequilibrium quantum and thermal fluctuations gives the nonequilibrium
fluctuation-dissipation theorem \cite{general}
\begin{eqnarray}
v(t,t^{\prime})=\int_{0}^t\!\!d\tau_1\!\!\int_{0}^{t^{\prime}}\!\!\!\!d\tau_2
~ u(t,\tau_1) {\widetilde g}(\tau_1,\tau_2)u^{\ast}(t^{\prime},\tau_2),
\label{vtb}
\end{eqnarray}
where ${\widetilde g}(\tau_1,\tau_2)\!=\!\int_0^{\infty} d\omega  J(\omega) {\bar n}(\omega,T) e^{-i\omega(\tau_1-\tau_2)}$
and ${\bar n}(\omega,T)=\frac{1}{e^{\hbar \omega / k_B T}-1}$ is the initial particle number distribution
of the environment.

The exact analytic solution of the integro-differential equation (\ref{ide}) is recently given in \cite{general}
as $u(t,t_0)=\int_{-\infty}^{\infty} d\omega {\mathcal D}(\omega) \exp \{-i\omega(t-t_0) \}$ with
\begin{eqnarray}
{\mathcal D} (\omega) = {\mathcal D}_l(\omega) + {\mathcal D}_c(\omega),
\label{ut2}
\end{eqnarray}
where ${\mathcal D}_l(\omega)={\mathcal Z} \delta (\omega-\omega_b)$ is the contribution of a
dissipationless localized mode, and ${\mathcal D}_c(\omega)=J(\omega)/[\{\omega-\omega_0-\Delta(\omega)\}^2 + \pi^2 J^2(\omega)]$
is the continuous part of the spectra. Here $\Delta(\omega) = {\cal P}\int_0^\infty d\omega^{\prime} \frac{J(\omega^{\prime})}{\omega-\omega^{\prime}}$ is a principal-value integral. The localized mode frequency
$\omega_b$ is determined by the pole condition $\omega_b - \omega_0 - \Delta(\omega_b)=0$, and
${\mathcal Z}=\left[ 1 - \Sigma^{\prime}(\omega_b) \right]^{-1}$ corresponds to the residue at the pole, which gives the
amplitude of the localized mode. Here, $\Sigma(\omega \pm i 0^{+})=\int_0^\infty d\omega^{\prime}\frac{J(\omega^{\prime})}{\omega-\omega^{\prime} \pm i 0^{+}}=\Delta(\omega)\mp i \pi J(\omega)$ is the
self-energy correction induced by the system-environment coupling. We consider an Ohmic spectral density \cite{LeggettRMP}
$J(\omega) = \eta \omega \exp\left(-\omega/\omega_c \right)$, where $\eta$ is the coupling strength between the system
and the environment, and $\omega_c$ is the frequency cutoff of the environmental spectra. For this case, a localized mode
appears when the coupling strength $\eta$ exceeds some critical value $\eta_c=\omega_0/\omega_c$. With the above specification, the
exact second-order correlation function $g^{(2)}(t,t+\tau)$ can be calculated explicitly and exactly through Eq.~(\ref{g2}), where the
numerator $\langle a^{\dagger}(t) a^{\dagger}(t+\tau) a(t+\tau) a(t) \rangle$ is given by Eq.~(\ref{2order}) and the denominator is
determined through $\langle a^{\dagger}(t) a(t) \rangle=|u(t)|^2 \alpha + v(t)$.

\begin{figure}[h]
\includegraphics[width=9.0cm]{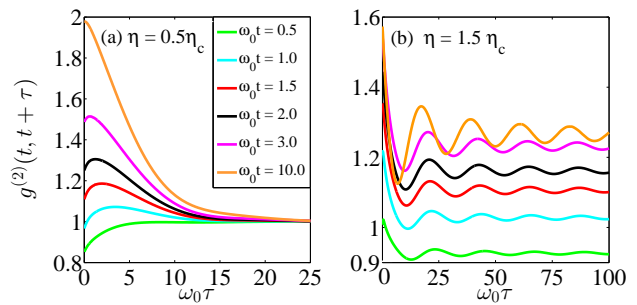}
\caption{\label{fg1} (Color online) Transient dynamics of $g^{(2)}(t,t+\tau)$ is shown
for two different system-environment coupling strengths: (a)  the weak coupling ($\eta=0.5\eta_c$)
and (b) the strong coupling ($\eta=1.5\eta_c$). Different curves represent different values of $t$ as shown by
the color legends. The other parameters are taken as $\omega_c = 5 \omega_0$, $k_{B}T = 2.0 \hbar\omega_0$,
and the system is considered to be in an initial photon number state $|n\rangle=|5\rangle$.}
\end{figure}
In Fig.~\ref{fg1}, we show the transient dynamics of the second-order correlation $g^{(2)}(t,t+\tau)$ as a
function of delay-time $\tau$ for different transient time $t$ for a given initial environment temperature,
$k_{B}T = 2.0 \hbar\omega_0$. We also let the system be initially in a Fock state \cite{Martinis,Haroche11} with an
arbitrary photon number $|n_0\rangle$, and consider the two cases of the system-environment coupling strength,
$\eta=0.5\eta_c$ (a weak coupling) and $\eta=1.5\eta_c$ (a strong coupling). Different curves in Fig.~\ref{fg1}
represent different transient time $t$. When the coupling strength is weak (see Figs.~\ref{fg1}a), we observe
photon antibunching in the short delay-time regime ($\omega_0\tau < 15$). For this short-time transient regime,
the magnitude of $g^{(2)}(t,t+\tau)$ rises with increasing the delay-time $\tau$, hence the photons show an
antibunching tendency. The antibunching gradually disappears at later time ($\omega_0t > 2.0$, for example).
Then $g^{(2)}(t,t+\tau)$ shows a monotonous decay with increasing $\tau$, which corresponds to the bunching
statistics. Hence, by measuring the correlation at different transient time $t$, one can have a smooth transition from
antibunching to bunching statistics. The second-order correlation function $g^{(2)}(t,t+\tau)$ approaches asymptotically
to unity in the long delay-time limit ($\omega_0 \tau > 20$), and the results also become independent of $t$
(the steady-state solution). Physically, this is intuitive from the density matrix evolution of the optical field. For the weak
coupling ($\eta < \eta_c$), the initial Fock state will always evolve to the following steady-state at thermal equilibrium
\cite{BreakBE}
\begin{eqnarray}
\label{dmatrix1}
\rho(t_s) = \sum_{n=0}^{\infty} \frac{[v(t_s)]^n}{[1+v(t_s)]^{n+1}} |n \rangle \langle n|,
\end{eqnarray}
which is solely determined by the steady-state value $v(t_s)={\bar n}(\omega_0,T)$ as the steady-state value of
$u(t_s) \rightarrow 0$. For a thermal field given by Eq.~(\ref{dmatrix1}), the steady-state correlation $g^{(2)}_{ss}(\tau)$
decays monotonically with increasing $\tau$, showing the familiar bunching statistics \cite{Scully97} with
$g^{(2)}(0) \rightarrow 2$ and $g^{(2)}(\infty) \rightarrow 1$. This result is consistent with the weak coupling steady-state
limit of $g^{(2)}$ given by Eq.~(\ref{g2M}).

However, for the strong system-environment coupling ($\eta > \eta_c$), the transient dynamics of $g^{(2)}(t,t+\tau)$
is significantly different (see Fig.~\ref{fg1}b). In this case, the magnitude of $g^{(2)}(t,t+\tau)$ decays first with increasing
the delay-time $\tau$. This corresponds to the bunching statistics. Then $g^{(2)}(t,t+\tau)$ starts rising with increasing the
delay-time $\tau$ and oscillate in $\tau$ (see Fig.~\ref{fg1}b). This result is very different from the result obtained in the
weak-coupling case, where $g^{(2)}(t,t+\tau)$ exhibits the antibunching statistics in the short-$\tau$ regime, as discussed above.
Physically, such bunching-antibunching oscillation is a manifestation of non-Markovian dynamics of the optical field
characterized by the reduced or enhanced correlation (\ref{2order}), originating from a localized mode contribution
of $u(t,t_0)$ given in Eq.~(\ref{ut2}). In fact, the two-time correlation function (\ref{2order}) correlates a past event with its
future providing useful information about the system-environment back-action memory effect. It was shown \cite{general}
that the non-Markovian dynamics of an open quantum system is fully characterized by the two-time correlation functions
$u(t,t_0)$ and $v(t,t^{\prime})$, the two-time correlation function is recently used to define a measure of
non-Markovianity \cite{ali2015}.

\begin{figure}[t]
\includegraphics[width=9.0cm]{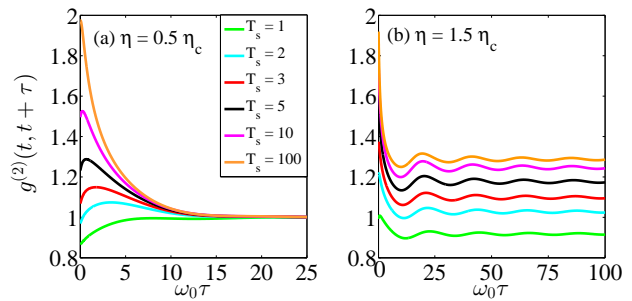}
\caption{\label{fg2} (Color online) Effect of environmental temperature on the transient dynamics of
$g^{(2)}(t,t+\tau)$ for two different system-environment coupling strengths: (a) weak coupling ($\eta=0.5\eta_c$)
and (b) strong coupling ($\eta=1.5\eta_c$). Different curves represent different values of $T_s$ shown by the color legends.
The other parameters are taken as $\omega_c = 5 \omega_0$, $\omega_0t=1$, and the system is considered to be in an initial
photon number state $|n\rangle=|5\rangle$.}
\end{figure}
In Fig.~\ref{fg2}, we show how the transient dynamics of $g^{(2)}(t,t+\tau)$ depends on the initial environment temperature $T$.
We consider again a weak-coupling case $\eta < \eta_c$ (see Fig.~\ref{fg2}a) and a strong-coupling case $\eta > \eta_c$
(see Figs.~\ref{fg2}b). The temperature dependence of $g^{(2)}(t,t+\tau)$ comes from the non-equilibrium fluctuation-dissipation
theorem through the initial particle number distribution ${\bar n}(\omega,T)$, given by Eq.~(\ref{vtb}). Different curves in Fig.~\ref{fg2}
represent different initial environment temperatures as shown by the color legends of each plot, where $T_s$ is defined as a dimensionless
temperature $T_s=k_{B}T/\hbar\omega_0$. When the coupling strength is weak ($\eta < \eta_c$), we observe photon
antibunching for low temperatures ($T_s < 5$) in the short delay-time regime ($\omega_0\tau < 10$). Here, a rising magnitude
of $g^{(2)}(t,t+\tau)$ is observed with increasing delay-time $\tau$ (see Fig.~\ref{fg2}a). The transient antibunching effect is
gradually suppressed when the initial environment temperature is increased. For a high initial environment temperature ($T_s > 5$),
the transient dynamics of $g^{(2)}(t,t+\tau)$ shows a monotonous decay of magnitude with $\tau$, manifesting a photon bunching
statistics. Figure~\ref{fg2}a essentially manifests the transition from antibunching to bunching photon statistics through the initial
environment temperature dependence of the correlation $g^{(2)}(t,t+\tau)$.

For the strong coupling case ($\eta > \eta_c$), $g^{(2)}(t,t+\tau)$ shows again  a short-$\tau$ oscillatory behavior
due to the non-Markovian dynamics from the localized mode contribution (see Fig.~\ref{fg2}b). This bunching-antibunching
oscillation in the strong-coupling regime becomes more visible as the initial environment temperature becomes higher and higher.
In this case, we also find that the correlation function $g^{(2)}(t,t+\tau)$ will saturate to various long-time steady-state values
(memory effect). This is unlike the case for the weak-coupling, where $g^{(2)}(t,t+\tau)$ asymptotically approaches to unity in the
steady-state limit.

\begin{figure}[h]
\includegraphics[width=5.5cm]{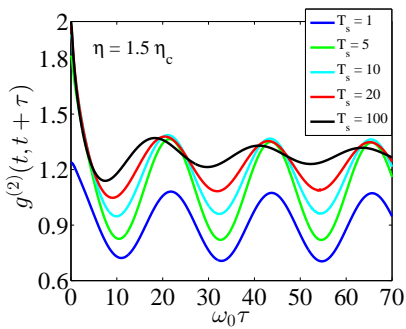}
\caption{\label{fg3} (Color online) We plot $g^{(2)}(t, t+ \tau)$ in the steady-state limit ($t \rightarrow \infty$)
for strong system-environment coupling. Different curves represent different values of temperature $T_s$ shown by
the color legends. The other parameters are taken as $\omega_c = 5 \omega_0$, and the system is considered to be in
an initial photon number state $|n_0\rangle=|5\rangle$.}
\end{figure}
Moreover, we observed an unusual nonequilibrium steady-state situation in the strong-coupling case. Usually, the steady-state
limit of $g^{(2)}(t, t+ \tau)$ discussed in the literature \cite{Scully97} is valid only for the weak-coupling regime, where the
two-time correlation function is calculated in the Markov limit.
In such a situation, one always obtains photon bunching statistics \cite{HBT1} in the steady-state limit, as discussed above,
and photon antibunching never shows up. In Fig.~\ref{fg3}, going beyond the weak coupling, we show the exact dynamics of
$g^{(2)}(t, t+ \tau)$ in the steady-state limit. Different solid curves in Fig.~\ref{fg3} demonstrate different initial
temperature-dependence of the environment. The steady-state limit of $g^{(2)}(t, t+ \tau)$ in the strong coupling shows the striking
photon bunching-antibunching oscillations. These oscillations survive longer (long-$\tau$ regime) at low temperatures,
but also persist even at a high temperature ($k_{B}T \!=\! 100 \hbar\omega_0$). Actually, for the strong-coupling case
($\eta > \eta_c$), the initial Fock state will evolve to a more complex steady-state \cite{Complexity} determined
by the steady-state values of both $u(t_s)= {\mathcal Z} \exp(-i \omega_b t_s)$ and $v(t_s)=\int_{0}^{\infty} d\omega
[ {\tilde {\mathcal D}}_l(\omega) + {\mathcal D}_c(\omega) ]  {\bar n}(\omega,T)$,
where ${\tilde {\mathcal D}}_l(\omega)=J(\omega){\mathcal Z}^2/(\omega-\omega_b)^2$:
\begin{eqnarray}
\label{dmatrix2}
\rho(t) = \sum_{n=0}^{\infty} p_n(t_s) |n \rangle \langle n|,
\end{eqnarray}
where
\begin{eqnarray}
\nonumber
\label{pnt}
p_n(t_s) &=& \frac{[v(t_s)]^n}{[1+v(t_s)]^{n+1}}[1-\Omega(t_s)]^{n_0} \\
\nonumber
&&{} \times \sum_{k=0}^{\textrm{min}\{n_0,n\}}
\left( \begin{array}{c}
n_0 \\
k
\end{array} \right)
\left( \begin{array}{c}
n \\
k
\end{array} \right)
\left[ \frac{1}{v(t_s)} \frac{\Omega (t_s)}{1-\Omega (t_s)} \right]^k,
\end{eqnarray}
and $\Omega (t_s) = |u(t_s)|^2/[1+v(t_s)]$. This is a nonequilibrium state that always depends on the initial state $|n_0\rangle$.
In other words, the system cannot approach to a thermal equilibrium state, due to the existence of the localized mode, as
shown in \cite{Mahan,Complexity,anderson}.

In conclusion, we have shown the exact transient dynamics of photon statistics for an optical
field coupled to a general non-Markovian environment. We observe a nontrivial transition
from antibunching to bunching statistics in the transient regime when the field  interacts {\it weakly}
with the environment. For the {\it strong} system-environment coupling, we find an interesting nonequilibrium
oscillatory dynamics between the photon bunching and the antibunching statistics that persists for arbitrary
initial temperature of the environment. Because experimentally one can prepare the system in a Fock state
\cite{Martinis,Haroche11}, the nontrivial nonequilibrium dynamics of the photon statistics discovered in this
work can be experimentally measured.

\begin{acknowledgments}
This work is supported by the Ministry of Science and Technology of Taiwan under Contract No.
NSC-102-2112-M-006-016-MY3 and The National Center for Theoretical Sciences. It is
also supported in part by the Headquarters of University Advancement at the National
Cheng Kung University, which is sponsored by the Ministry of Education, Taiwan, ROC.
\end{acknowledgments}

\end{document}